# ON THE INHERENT FIGURE OF MERIT OF THERMOELECTRIC COMPOSITES

*A.A. Snarskii[1,2], G.V.Adzhigai[2], I.V.Bezsudnov[3]*
*([1,2] Institute of Thermoelectricity, Chernivtsi, Ukraine;*
*[2] "KPI" National Technical University of Ukraine, Kyiv, Ukraine;*
*[3] "Nauka-Service" Ltd, Moscow, Russia)*

*The paper introduces and analyzes a new characteristic of thermoelectric composites – their intrinsic figure of merit characterizing the influence of thermoelectric phenomena on the effective properties of composites and governing thermal into electric energy conversion within a composite.*

**Introduction**

The homogeneous semiconductor material is characterized by three material constants – electric conductivity $\sigma$, thermal conductivity $\kappa$ and differential thermopower $\alpha$. When describing energy processes (efficiency, cooling power, etc.) the above three constants make possible a natural (and unique) way of introducing a dimensionless temperature

$$\tilde{T} = \frac{\sigma \alpha^2}{\kappa} T. \qquad (1)$$

This dimensionless temperature and, respectively, the combination $\sigma\alpha^2/\kappa = Z$, first introduced by A. Ioffe (see, for example, [1]) and referred to as figure of merit, Ioffe's number, is decisive for the efficiency. The higher $\tilde{T} = ZT$, the higher the thermocouple efficiency.

The material used is often a composite. Hereinafter we shall consider two-phase macroscopically inhomogeneous media. The properties of each of the phases in this case can be characterized by their local constants $\sigma_i$, $\kappa_i$, $\alpha_i$ ($i = 1, 2$), each of the phases being considered isotropic. Randomly inhomogeneous media are generally characterized by effective kinetic coefficients (EKC) $\sigma_e$, $\kappa_e$, $\alpha_e$, that by definition interrelate volume average thermoelectric forces and flows [2, 3]. For thermoelectric media it means that if the following holds true locally

$$\mathbf{j} = \sigma \mathbf{E} - \sigma\alpha\nabla T,$$
$$\mathbf{q}/T = -\kappa\left(1 + \frac{\sigma\alpha^2}{\kappa}T\right)\nabla T + \sigma\alpha\mathbf{E}, \qquad (2)$$

where $\mathbf{j}$ and $\mathbf{q}$ are the densities of electric current and heat flow, and $\nabla T$ is temperature gradient, then for volume average current $\langle\mathbf{j}\rangle$, field $\langle\mathbf{E}\rangle$, heat flow $\langle\mathbf{q}\rangle$ and temperature gradient $\langle\nabla T\rangle$

$$\langle\mathbf{j}\rangle = \sigma_e\langle\mathbf{E}\rangle - \sigma_e\alpha_e\langle\nabla T\rangle,$$
$$\langle\mathbf{q}/T\rangle = -\left(1 + \frac{\sigma_e\alpha_e^2}{\kappa_e}T\right)\kappa_e\langle\nabla T\rangle + \sigma_e\alpha_e\langle\mathbf{E}\rangle. \qquad (3)$$

The EKC $\sigma_e$, $\kappa_e$ и $\alpha_e$ are functionals of $\mathbf{E}(\mathbf{r})$ and $\nabla T(\mathbf{r})$ and intricately depend on the local values of $\sigma_i$, $\kappa_i$, $\alpha_i$ and phase concentrations. Their calculation methods are described in a brief review [3].

While for the homogeneous material the figure of merit $Z = \sigma\alpha^2/\kappa$ is **determined** unambiguously, the two-phase composites lack this unambiguity and there are many different variants for making "dimensionless" temperature and, accordingly, "figures of merit": $\sigma_e\alpha_e^2/\kappa_e$, $\sigma_1\alpha_1^2/\kappa_1$, $\sigma_2\alpha_2^2/\kappa_2$, $(\sigma_1 + \sigma_2)(\alpha_1 + \alpha_2)^2 / (\kappa_1 + \kappa_2)$, ... . As will be shown below, in addition to phase figure of merit $Z_i = \sigma_i \alpha_i^2 / \kappa_i$ and effective figure of merit $Z_e$

$$Z_e = \sigma_e \alpha_e^2 / \kappa_e, \qquad (4)$$

there is at least one more important combination with figure of merit dimensions that has a clear physical meaning. If $Z_e$ assigns the thermocouple efficiency, i.e. **determines** conversion of heat from the external source to electricity, also transferred to the external circuit, there is a value that **determines** conversion of thermal into electric energy dissipating within a composite. It would appear natural that such "figure of merit" should be called the inherent figure of merit.

Note that even in the homogeneous case *ZT* acts in two capacities**.** On the one hand, *ZT* **determines** the efficiency, on the other hand, it **renormalizes** thermal conductivity. While a coefficient appearing in the expression for heat flow density (2) with temperature gradient and **determining** heat flow generated by temperature gradient at $ZT \ll 1$ is a "conventional" thermal conductivity $\kappa$, at $ZT > 1$ this coefficient is already by a factor of *ZT* greater.

**Inherent figure of merit of thermoelectric materials**

As long as in this paper we are not aiming at calculation of specific devices, our objective is to show principal existence and physical meaning of the new introduced value – latent figure of merit $\tilde{Z}$, let us refer to the case of simpler two-dimensional randomly inhomogeneous medium. In a number of cases it has a precise solution for EKC $\sigma_e$, $\kappa_e$, $\alpha_e$ [4, 5, 6]. This solution takes place for the so-called self-dual media [4] (see also [7]), including two-dimensional randomly inhomogeneous media at flow threshold $p_c$, which in a two-dimensional case is equal to $p_c = 1/2$. In the case of self-dual media [5, 6] the EKC are of the form

$$\sigma_e = \sqrt{\sigma_1 \sigma_2} \frac{\sqrt{\sigma_1 \kappa_2} + \sqrt{\sigma_2 \kappa_1}}{\sqrt{\left(\sqrt{\sigma_1 \kappa_2} + \sqrt{\sigma_2 \kappa_1}\right)^2 + T \sigma_1 \sigma_2 (\alpha_1 - \alpha_2)}}, \qquad (5)$$

$$\kappa_e = \sqrt{\kappa_1 \kappa_2} \frac{\sqrt{\sigma_1 \sigma_2}}{\sigma_e},$$

$$\alpha_e = \frac{\alpha_1 \sqrt{\sigma_1 \kappa_2} + \alpha_2 \sqrt{\sigma_2 \kappa_1}}{\sqrt{\sigma_1 \kappa_2} + \sqrt{\sigma_2 \kappa_1}}. \qquad (6)$$

Precise (valid for arbitrarily large inhomogeneity) expressions for $\sigma_e$, $\kappa_e$, $\alpha_e$ in the entire concentration range are unknown (and hardly possible). One of the most successful approximations is that of a self-consistent field [8] (see also the books [9, 10]). In the case when corrections $\kappa$ $\sigma_e$ and $\kappa_e$ due to thermoelectric phenomena are small, according to [11] (see also [6]) $\alpha_e = \frac{\langle \alpha \sigma / \Delta \rangle}{\langle \sigma / \Delta \rangle}$, $\Delta = (\sigma_{lin}^e + \sigma)(\kappa_{lin}^e + \kappa)$,

(7)

$$\left\langle \frac{\sigma_{lin}^e - \sigma}{\sigma_{lin}^e + \sigma} \right\rangle = 0, \quad \left\langle \frac{\kappa_{lin}^e - \kappa}{\kappa_{lin}^e + \kappa} \right\rangle = 0, \qquad (8)$$

where the subscript denotes that the EKC are taken as a linear approximation, ignoring the influence of thermoelectric phenomena on the effective electric conductivity and thermal conductivity.

In an explicit form, for a two-phase case of (7-8)

$$\alpha_{lin}^{e} = \frac{p\sigma_{1}\alpha_{1}\left(\sigma_{lin}^{e}+\sigma_{2}\right)\left(\kappa_{lin}^{e}+\kappa_{2}\right)+(1-p)\sigma_{2}\alpha_{2}\left(\sigma_{lin}^{e}+\sigma_{1}\right)\left(\kappa_{lin}^{e}+\kappa_{1}\right)}{p\sigma_{1}\left(\sigma_{lin}^{e}+\sigma_{2}\right)\left(\kappa_{lin}^{e}+\kappa_{2}\right)+(1-p)\sigma_{2}\left(\sigma_{lin}^{e}+\sigma_{1}\right)\left(\kappa_{lin}^{e}+\kappa_{1}\right)}, \qquad (9)$$

$$\sigma_{lin}^{e} = \left(p-\frac{1}{2}\right)(\sigma_{1}-\sigma_{2})+\left[\left(p-\frac{1}{2}\right)^{2}(\sigma_{1}-\sigma_{2})^{2}+\sigma_{1}\sigma_{2}\right]^{\frac{1}{2}}, \qquad (10)$$

$$\kappa_{lin}^{e} = \left(p-\frac{1}{2}\right)(\kappa_{1}-\kappa_{2})+\left[\left(p-\frac{1}{2}\right)^{2}(\kappa_{1}-\kappa_{2})^{2}+\kappa_{1}\kappa_{2}\right]^{\frac{1}{2}}. \qquad (11)$$

For $p = p_{c} = 1/2$ from (10) and (11) it follows

$$\sigma_{lin}^{e} = \sqrt{\sigma_{1}\sigma_{2}}, \quad \kappa_{lin}^{e} = \sqrt{\kappa_{1}\kappa_{2}}. \qquad (12)$$

Equations to **determine** the EKC within the approximation of a self-consistent field in the general case with regard to the influence of thermoelectric phenomena on electric and thermal conductivity can be solved only numerically. According to [6] these equations are of the form

$$pA_{1}+(1-p)A_{2}=1,\ pB_{1}+(1-p)B_{2}=0,\ pC_{1}+(1-p)C_{2}=0,\ pD_{1}+(1-p)D_{2}=1, \qquad (13)$$

where

$$A_{i} = \frac{2}{\Delta_{i}}\left[\sigma_{e}\left(\chi_{e}+\chi_{i}\right)-\gamma_{e}\left(\gamma_{e}+\gamma_{i}\right)\right],\ B_{i} = \frac{2}{\Delta_{i}}\left(\gamma_{e}\chi_{i}-\chi_{e}\gamma_{i}\right),$$

$$C_{i} = \frac{2}{\Delta_{i}}\left(\gamma_{e}\sigma_{i}-\sigma_{e}\gamma_{i}\right),\ D_{i} = \frac{2}{\Delta_{i}}\left[\chi_{e}\left(\sigma_{e}+\sigma_{i}\right)-\gamma_{e}\left(\gamma_{e}+\gamma_{i}\right)\right],$$

$$\Delta_{i} = \left(\sigma_{e}+\sigma_{i}\right)\left(\chi_{e}+\chi_{i}\right)-\left(\gamma_{e}+\gamma_{i}\right)^{2},\ \gamma=\sigma\alpha,\ \chi=T^{-1}\kappa+\sigma\alpha^{2}.$$

As an illustration, with numerical calculations we shall consider the first composite phase to be metal and the second - semiconductor: $\sigma_{1} \gg \sigma_{2}$, $\kappa_{1} > \kappa_{2}$, $\alpha_{2} \gg \alpha_{1}$. For certainty, let us assume $\alpha_{1}$=2.4·10$^{-6}$ V/K, $\kappa_{1}$=2.6·10 W/(m·K), $\kappa_{2}$=2.6 W/(m·K), $\sigma_{1}$=2.6·10$^{2}$ (Ohm·m)$^{-1}$, $\sigma_{2}$=2.6 (Ohm·m)$^{-1}$, $T$=300 K. ThermoEMF of the second phase $\alpha_{2}$ will be changed in the range (2.2·10$^{-4}$ 900·2.2·10$^{-4}$) V/K. Generally speaking, the right boundary of $\alpha_{2}$ is strongly overrated as compared to currently available materials. With such value of $\alpha_{2}$=900·2.2·10$^{-4}$ V/K and with chosen $\sigma_{2}$ and $\kappa_{2}$, $Z_{2}T \approx 12$, which is much in excess of known values. However, in the first place, $ZT$ of new materials is constantly growing, in the second place, large values of $\alpha_{2}$ and $Z_{2}T$ allow a better perception of qualitative differences between the linear and nonlinear cases.

Fig. 1 shows a concentration dependence of $Z_{e}T$ obtained within the linear and nonlinear approximations. It is evident that account of nonlinearity leads to reduction of figure of merit, and the higher $\alpha_{2}$, i.e. the higher $Z_{2}T$, the greater this reduction. It should be noted that the difference between figures of merit in the nonlinear and linear approximations $Z_{nlin}^{e} - Z_{lin}^{e}$ with $p \to 0,1$ tends to zero, and for a homogeneous material it is not to be observed at all. The respective ratio between the effective electric conductivity and thermal conductivity has a similar behaviour – Fig. 2.

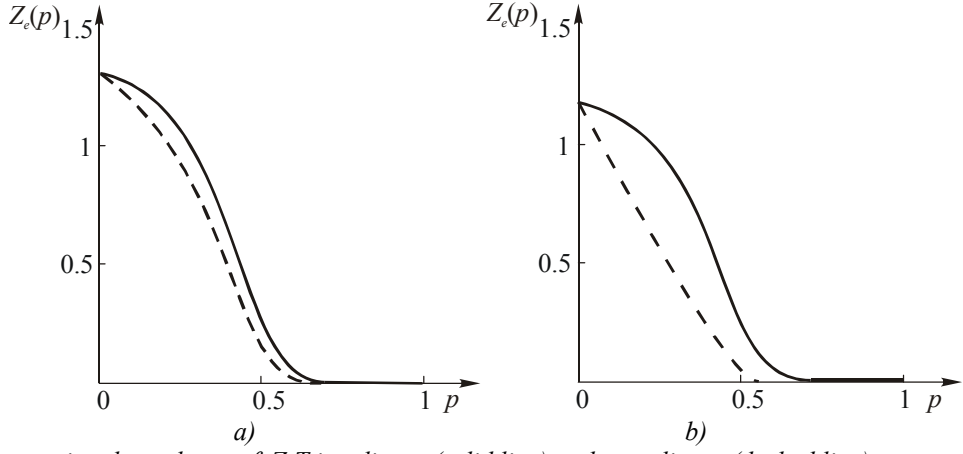

Fig. 1. Concentration dependence of $Z_eT$ in a linear (solid line) and a nonlinear (dashed line) case:
a) $\alpha_2=0.066$ V/K; b) $\alpha_2=0.198$ V/K.

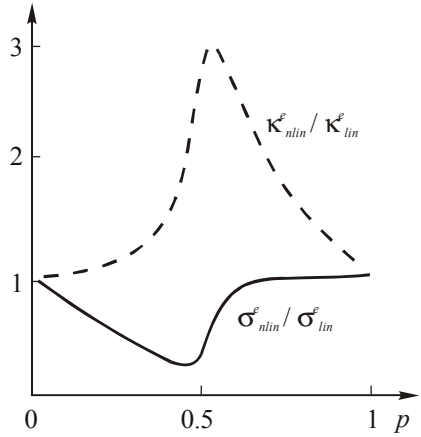

Fig. 2. The influence of inhomogeneity on the effective electric and thermal conductivity.
Parameter values have been taken as follows: $\alpha_1=2.4 \cdot 10^{-6}$ V/K, $\alpha_2=19.8 \cdot 10^{-2}$ V/K,
$\kappa_1=2.6 \cdot 10$ W/(m·K), $\kappa_2=2.6$ W/(m·K), $\sigma_1=2.6 \cdot 10^2$ (Ohm·m)$^{-1}$, $\sigma_2=2.6$ (Ohm·m)$^{-1}$, $T=300$ K.

With $p=p_c=1/2$, when analytical expressions for EKC are known not only in a linear, but also in a nonlinear case [6] (5-6), the influence of nonlinearity can be distinguished explicitly. Indeed, one can readily see that (5) can be rewritten as (see.(12))

$$\sigma^e_{nlin} = \frac{\sigma^e_{lin}}{\sqrt{1+\tilde{Z}T}}, \quad \kappa^e_{nlin} = \kappa^e_{lin}\sqrt{1+\tilde{Z}T}, \qquad (14)$$

where

$$\tilde{Z}T = \frac{\sqrt{\sigma_1\sigma_2}}{\sqrt{\kappa_1\kappa_2}} \cdot \frac{(\alpha_1-\alpha_2)^2}{\left(\sqrt{\frac{\sigma_2}{\sigma_1}}\bigg/\sqrt{\frac{\kappa_2}{\kappa_1}}+\sqrt{\frac{\kappa_2}{\kappa_1}}\bigg/\sqrt{\frac{\sigma_2}{\sigma_1}}\right)^2} T. \qquad (15)$$

Here $\tilde{Z}T$ is a dimensionless parameter hereinafter referred to as the inherent figure of merit of composite for a two-phase case at the flow threshold.

For a material that is homogeneous, at least in thermoelectric properties ($\alpha_1=\alpha_2$), the inherent figure of merit $\tilde{Z}T=0$. At $\tilde{Z}T \ll 1$, $\sigma^e_{nlin} \to \sigma^e_{lin}$, $\kappa^e_{nlin} \to \kappa^e_{lin}$. The increase in the inherent figure of merit $\tilde{Z}T$ results in the reduction of effective electric conductivity and the increase in effective thermal conductivity (Fig. 2), leading to a decrease in effective figure of merit $\tilde{Z}^e_{nlin} \sim \sigma^e_{nlin}/\kappa^e_{nlin}$.

Unfortunately, at $p \neq 1/2$ the complexity of nonlinear equations (13) does not allow obtaining $\sigma^e_{nlin}$ and $\kappa^e_{nlin}$ in the explicit form and thus distinguishing $\tilde{Z}(p)T$. Numerical solutions of system (13) in the absence of analytical form of dependence of $\sigma^e_{nlin}$ and $\kappa^e_{nlin}$ on $\tilde{Z}T$ do not allow obtaining concentration dependence $\tilde{Z} = \tilde{Z}(p)$ even numerically. However, in a partial case, namely for composites having electric and thermal phase conductivities that meet the law of Wiedemann-Franz

$$\frac{\sigma_1}{\kappa_1} = \frac{\sigma_2}{\kappa_2}, \qquad (16)$$

one can numerically obtain $\tilde{Z} = \tilde{Z}(p)$, or **in more conservative terms**, a function that is monotonously dependent on $\tilde{Z}(p)$.

To distinguish from $\sigma^e_{nlin}$ and $\kappa^e_{nlin}$ the inherent figure of merit $\tilde{Z}$ means to distinguish that part which appears with account of nonlinearity, and this part should be identical for $\sigma^e_{nlin}$ and $\kappa^e_{nlin}$. Exactly to this part the inherent figure of merit $\tilde{Z}$ is proportional. Let us introduce the notations

$$S(p) = \frac{\sigma^e_{lin}(p)}{\sigma^e_{nlin}(p)}, \quad K(p) = \frac{\kappa^e_{nlin}(p)}{\kappa^e_{lin}(p)}. \qquad (17)$$

In the case of $p = 1/2$

$$S(1/2) = K(1/2) = \sqrt{1 + \tilde{Z}T}, \qquad (18)$$

the following holds true

$$S(1/2)/K(1/2) = 1. \qquad (19)$$

Suppose that at $p \neq 1/2$ there is a certain function identical for $\sigma^e_{nlin}$ and $\kappa^e_{nlin}$ and **determining** the influence of nonlinearity. Then using $S(p)$ and $K(p)$ one can construct a certain combination similar to (19), that at any $p$ will be identical to unity and at $p = 1/2$ will coincide with (19). As is shown by numerical calculation to an accuracy of $2 \cdot 10^{-6}$, such combination is $S(p)/K(1-p)$. With fulfillment of (16) the following equation holds true

$$S(p)/K(1-p) = 1. \qquad (20)$$

Then the value directly related to the inherent figure of merit $\tilde{Z}(p)$, can be **determined** as (see. (18)) a certain function of $a(p) = S(p)K(1-p)$ and $b(p) = S(1-p)K(p)$

$$\tilde{Z}(p)T = M(a(p), b(p)) - 1. \qquad (21)$$

This function $M(a,b)$ should satisfy the following natural conditions: be 1) continuous, 2) symmetric $M(a,b) = M(b,a)$, 3) the average of identical functions $a(p) = b(p)$ should be equal to their total value $M(a,a) = a$. These requirements are nothing but the so-called axioms of the average first formulated by A.I.Kholmogorov [12]. In his work it was shown that if function $M(a,b)$ satisfies these axioms (for more than two variables one more axiom must be met), then $M(a,b)$ is called the average value of $a$, $b$ and can be written in the most general form as

$$M(a,b) = \psi\left(\frac{\varphi(a) + \varphi(b)}{2}\right). \qquad (22)$$

where $\varphi$ is a continuous, strictly monotonous function, and $\psi$ is the inverse to it.

All known types of averages, such as arithmetic mean, quadratic mean, geometric mean, harmonic mean, etc are a partial case (22).

Thus, $M(a(p), b(p))$ (21) can be written in the form of various averages and is not determined unambiguously

$$M_1(p) = \frac{S(p)K(1-p) + S(1-p)K(p)}{2}, \quad (23)$$

$$M_2(p) = 2\frac{S(p)K(1-p)S(1-p)K(p)}{S(p)K(1-p) + S(1-p)K(p)}, \quad (24)$$

$$M_3(p) = \sqrt{S(p)K(1-p)S(1-p)K(p)}, \ldots \quad (25)$$

In all these cases $\tilde{Z}(p=1/2)$ from (21) coincides with its value $\tilde{Z}$ from (15). Fig. 3 shows a concentration dependence of $\tilde{Z}(p)T$ from (21) for three types of function $M$ (23), (24) and (25). As can be seen from Fig.3, the difference between the concentration dependences $\tilde{Z}(p)T$ for selected numerical values of local coefficients is slight.

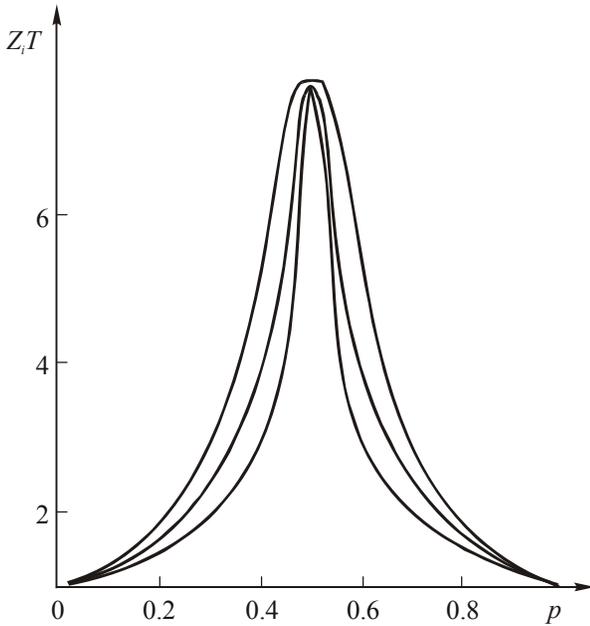

Fig. 3. Concentration dependence of the inherent figure of merit, related to average $M_i$ (top-down in order) (23), (24), (25).

Note that among the axioms of average the axiom of symmetry $M(a,b)$ is, in our view, natural for the **determination** of the inherent figure of merit. When this axiom is abandoned, the class of functions for $\tilde{Z}(p)T$ from (21) is expanded considerably.

It is of interest to study the inherent figure of merit $\tilde{Z}$ in a three-dimensional case. It is also interesting to consider $\tilde{Z}$ beyond the approximation limits of a self-consistent field (even in a two-dimensional case). Apparently, it is possible only with numerical simulation of thermoelectric processes in the inhomogeneous media, with the use of special application packages, and will be the subject of a separate publication.

**Conclusions**

1. In addition to figure of merit (Ioffe's number) of composite material $Z_e = \sigma_e \alpha_e^2 / \kappa_e$, characterizing «external» thermal into electric energy conversion (for example, thermocouple efficiency) there is yet another characteristic, i.e. the inherent figure of merit $\tilde{Z}$, **governing** energy conversion within a composite.

2. The inherent figure of merit $\tilde{Z}$ **determines renormalization** of composite effective thermal and electric conductivity due to thermoelectric phenomena and is a limiting factor for the effective figure of merit of composite material.